\documentclass[11pt,twoside]{article}
\usepackage{asp2006}
\usepackage{psfig}
\usepackage{epsf}
\usepackage{graphics}
\usepackage{lscape}
\begin{document}
\title{Large Eddy Simulation of Solar Photosphere Convection with
Realistic Physics}
\author{S. D. Ustyugov}
\affil{Keldysh Institute of Applied Mathematics, 4, Miusskaya sq., 
Moscow, Russia}

\begin{abstract}
Three-dimensional large eddy simulations of solar surface convection using realistic model
physics are conducted. The thermal structure of convective motions into the upper radiative 
layers of the photosphere, the range of convection cell sizes, and the penetration depths
of convection are investigated. A portion of the solar photosphere and the upper layers of the
convection zone, a region extending $60\times60$ Mm horizontally from 0 Mm down to 20 Mm below
the visible surface, is considered. We start from a realistic initial model of the Sun with an
equation of state and opacities of stellar matter. The equations of fully compressible radiation
hydrodynamics with dynamical viscosity and gravity are solved. We use:
1) a high order conservative TVD scheme for the hydrodynamics,
2) the diffusion approximation for the radiative transfer,
3) dynamical viscosity from subgrid scale modeling.
The simulations are conducted on a uniform horizontal grid of $600\times600$, with 168 nonuniformly 
spaced vertical grid points, on 144 processors with distributed memory multiprocessors
on supercomputer MVS-15000BM in the Computational Centre of the Russian Academy of Sciences.

\end{abstract}

\section{Introduction}
Convection near the solar surface has a strongly non-local and dynamical
character. Hence, numerical simulations provide useful information on
the spatial structures resulting from convection and help in constructing
consistent models of the physical processes underlying the observed
solar phenomena. We conduct an investigation of the temporal evolution and growth 
of convective modes on scales of mesogranulation and supergranulation in a 
three-dimensional computational box. In previous work by the author [\citet{ustyugs06}]
it was shown that collective motion of small convective cells of granulation expels 
weak magnetic field on the edges of cells at mesogranular scales. The average size of such cells 
is 15-20 Mm and the lifetime of order 8-10 solar hours. Simulation of solar photosphere 
convection [\citet{stein06}] in a computational domain of size 48 Mm in the horizontal 
plane and 20 Mm in depth showed that the sizes of convective cells increase with depth. 
The purpose of this work is to investigate the development and scales of convection 
in a region of size 60 Mm in the horizontal plane and 20 Mm in depth. 

\section{Numerical method}
The distribution of the main thermodynamic variables with radius was taken
from the Standard Solar Model [\citet{christ03}] with parameters
$(X,Z,\alpha)=(0.7385,0.0181,2.02)$, where $X$ and $Y$ are hydrogen
and helium abundances by mass, and $\alpha$ is the ratio of mixing
length to pressure scale height in the convection region. We used the OPAL opacity 
tables and the equation of state for solar matter [\citet{rogers96}].

The fully compressible nonideal hydrodynamics equations were solved:

\begin{equation}
\frac{\partial \rho}{\partial t}+ \nabla \cdot \rho \vec v = 0
\end{equation}

\begin{equation}
\frac{\partial \rho \vec v}{\partial t}+ \nabla \cdot \left[\rho \vec v \vec v
+ P \hat I \right]
= \rho \vec g + \nabla \cdot \hat \tau
\end{equation}

\begin{equation}
\frac{\partial E}{\partial t}+ \nabla \cdot \left[\vec v
\left(E + P \right) \right]= \nabla \cdot \left(\vec v \cdot \hat \tau \right) 
+ \rho \left(\vec g \cdot \vec v \right) + Q_{rad}
\end{equation}
$E=e+\rho v^2/2$ is the total energy, $Q_{rad}$ is the energy transferred by 
radiation and $\hat \tau$ is the viscous stress tensor. The influence of small scales 
on large scale flows was evaluated in terms of the viscous stress tensor and the rate 
of dissipation was defined from the buoyancy and shear production terms [\citet{canuto94}]. 
The evolution of all variables in time was found using an explicit TVD[Total Variation 
Diminishing] conservative difference scheme [\citet{yee90}]:
\begin{equation}
U^{n+1}_{i,j,k}=U^n_{i,j,k}-\Delta tL(U^n_{i,j,k}),
\end{equation}
where $\Delta t=t^{n+1}-t^n$ and the operator $L$ is
\begin{eqnarray}
L(U_{i,j,k})=\frac {\tilde F_{i+1/2,j,k}-\tilde F_{i-1/2,j,k}}{\Delta 
x_i}
+ \frac {\tilde G_{i,j+1/2,k}-\tilde G_{i,j-1/2,k}}{\Delta y_j}\nonumber \\
+ \frac {\tilde H_{i,j,k+1/2}-\tilde H_{i,j,k-1/2}}{\Delta 
z_k}+S_{i,j,k} \ .
\end{eqnarray}
The flux along each direction was defined by the local-characteristic method as 
\begin{equation}
\tilde F_{i+1/2,j,k}=\frac {1}{2}\left[F_{i,j,k}+F_{i+1,j,k}+
R_{i+1/2}W_{i+1/2}\right] \ .
\end{equation}
$R_{i+1/2}$ is the matrix whose columns are right eigenvectors of
$\partial F/\partial U$ evaluated as a generalized Roe average 
of $U_{i,j,k}$ and $U_{i+1,j,k}$ for real gases. $W_{i+1/2}$ is the matrix of numerical 
dissipation. The term $S_{i,j,k}$ represents the effect of gravitational forces and 
radiation. Time step integration is by third-order Runge-Kutta method [\citet{shu88}]. 
This scheme is second-order in space and time. Central differences were used for the 
viscous term, and the diffusion approximation was applied for the radiative term in 
the energy equation. 

We used a uniform grid in the $x$ and $y$ directions and a nonuniform grid in 
the vertical $(z)$ direction. Periodic boundary conditions were used in the horizontal 
directions and the top and bottom boundary conditions were chosen to be
$$v_{z,k}=-v_{z,k-1},v_{x,k}=v_{x,k-1},v_{y,k}=v_{y,k-1},dp/dz=\rho g_{z},p=p(\rho),e=const$$
that is, reflection for the $z$ component of velocity, outflow for the $x$ and $y$ components. 
Pressure and density were derived from the solution of the hydrostatic
equation, using the equation of state with a constant value of the internal energy.

\section{Results}

The numerical simulation of solar photosphere convection was conducted  
during 24 solar hours. On the upper boundary of the region, which corresponds
to optical depth $\tau = 1$ near the solar surface in the horizontal plane, 
the development of small-scale convection structure is clearly seen, with cells of
average size 1.5 Mm and lifetime of order 1-2 Min (Figure 1). At depth
1 Mm from the solar surface we observe the appearance of signs of large-scale
convective cells at the scale of mesogranulation (Figure 2). With increasing
depth the size of big convective cells grows and their number decreases.
At depth 2-5 Mm the average size of a cell is 10-15 Mm, whereas it is 15-20 Mm 
at depths greater than 10 Mm (Figure 3-6). Vertical motion on the edges
of convective cells has the maximum values of velocity and become supersonic at
cell vertices at depths 3-4 Mm. In the deepest part of the domain
the flow of material becomes slower and more laminar. 

By looking at the convective flow in a vertical plane we can see three regions, 
distinctly different in physical character (Figure 7-8). In the turbulent zone 
at depths 0-2 Mm, small-scale chaotic motion of granules due to the action 
of strong compressible effects leads to formation of downdrafts of cold blobs 
of material. There are places with strong vorticity motion where material moves 
from different directions to one point with the formation of more powerful 
downdrafts. In the transition zone at depths of 2-5 Mm we find large-scale flow, on 
the scales of mesogranulation. Here the downdrafts achieve a maximum velocity, 
on average, of about 4 km/sec with Mach number 1.5. Material moves downwards in the
form of narrow jets and upwards in wider regions with an average of size 10-15 Mm.
In the third part of the region, at greater depths, we obesrve some points where 
material reaches the bottom boundary with subsonic velocity. The distances between 
these points gives us the size of convective cells on the scale of supergranulation,
about 20 Mm.

We averaged the velocity components over 4 solar hours for one horizontal plane near 
the solar surface. We then introduced many Lagrange particles or corks for each cell, 
uniformly distributed across the plane at some moment of time.
After 4 solar hours of simulation we defined the places with the maximum number of 
corks. From Figure 9 it is clearly seen that large-scale divergent flows from the 
centers of big convective cells expel corks to their edges. Corks are located in 
places with converging flows (Figure 9) where we have positive values of the 
two-dimensional divergence of velocity and at points with strong vorticity
motion where we find large negative values of curl velocity. As seen from Figure 10 
there are convective cells with the size of supergranulation - about 20 Mm 
in diameter. Matter move away from the center of these cells  with average 
velocity 1-1.5 km/sec.

{\bf Acknoledgments.} I am grateful to Rudi Komm, National Solar Observatory and to 
NASA for financial support for my participation in NSO Workshop 24 at 
Sacramento Peak, New Mexico, USA.

\newpage

\begin{figure}[!ht]
  \plotfiddle{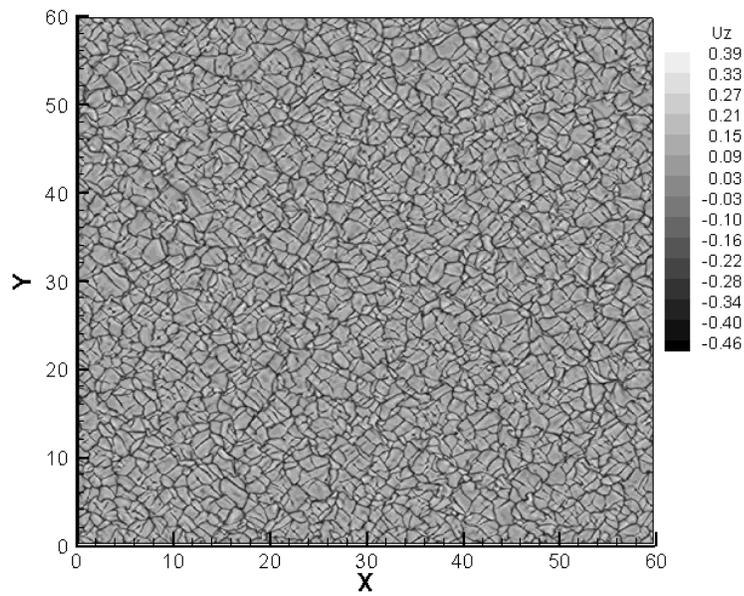}{50mm}{-90}{45}{45}{-200}{180}
  \vspace{2.5cm}
  \caption{\small Contours of the vertical component of velocity at depth 0 Mm . }
\end{figure}

\begin{figure}[!ht]
  \plotfiddle{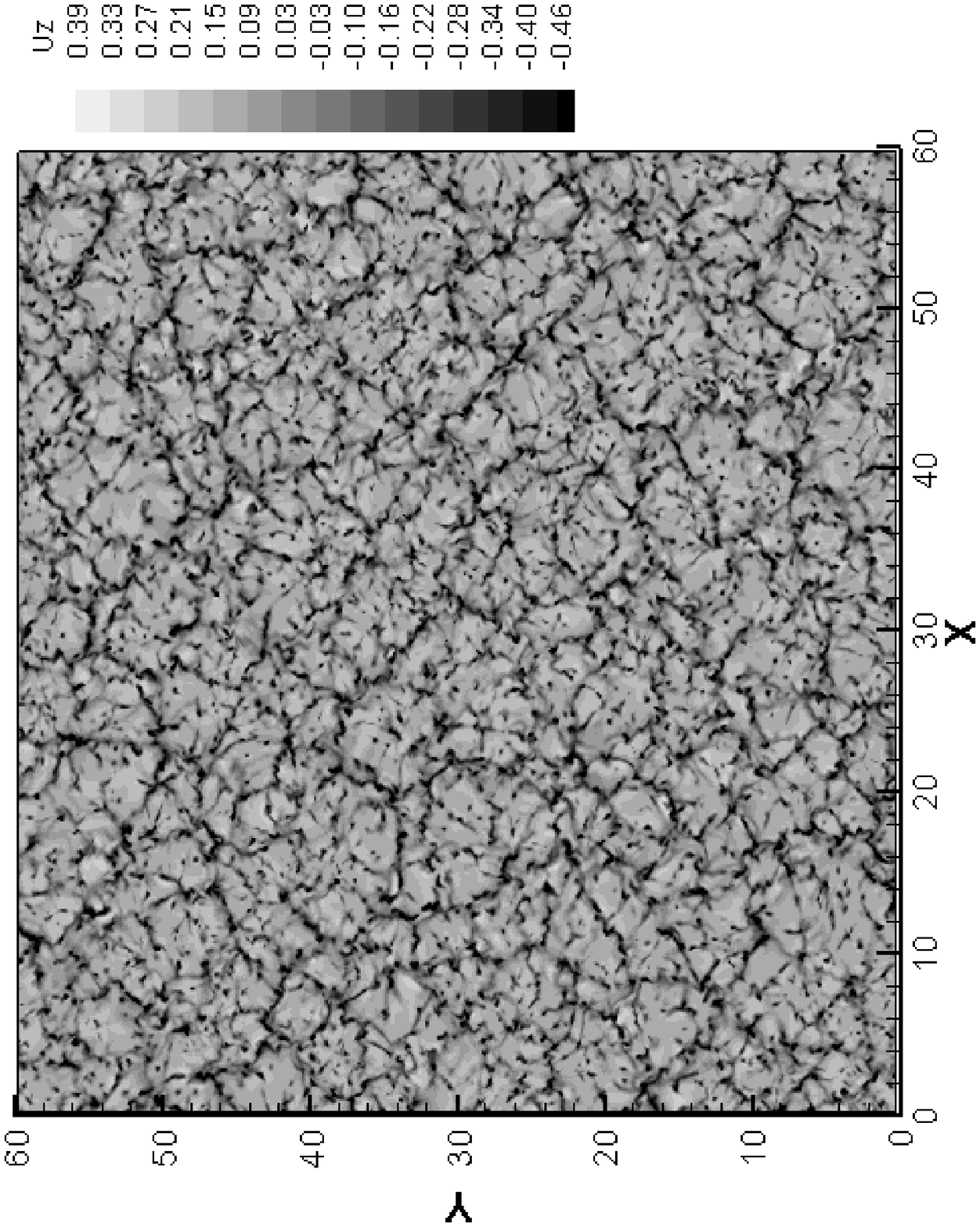}{50mm}{-90}{45}{45}{-200}{180}
  \vspace{2.5cm}
  \caption{\small Contours of the vertical component of velocity at depth 1 Mm. }
\end{figure}

\begin{figure}[!ht]
  \plotfiddle{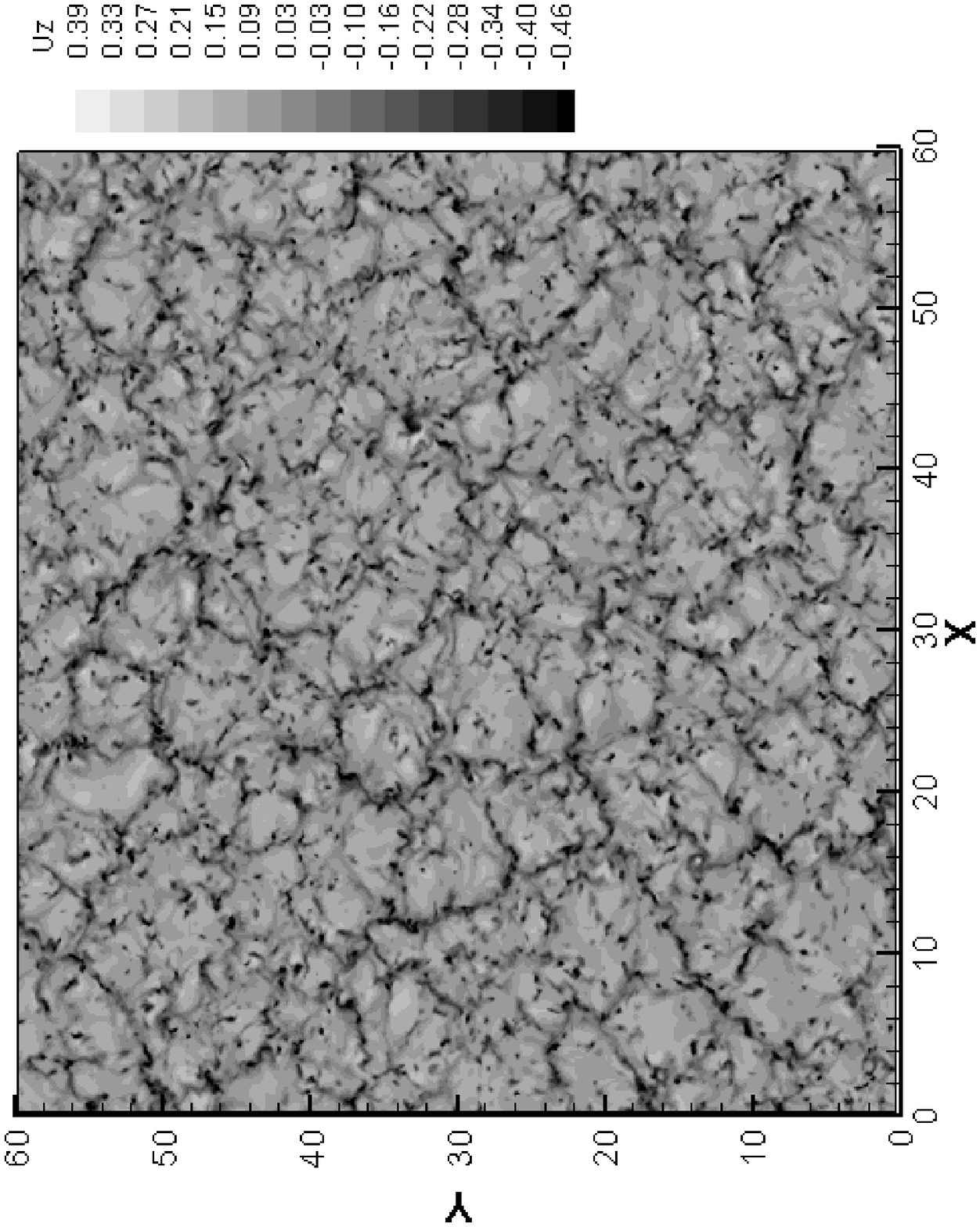}{50mm}{-90}{45}{45}{-200}{180}
  \vspace{2.5cm}
  \caption{\small Contours of the vertical component of velocity at depth 2 Mm. }
\end{figure}

\begin{figure}[!ht]
  \plotfiddle{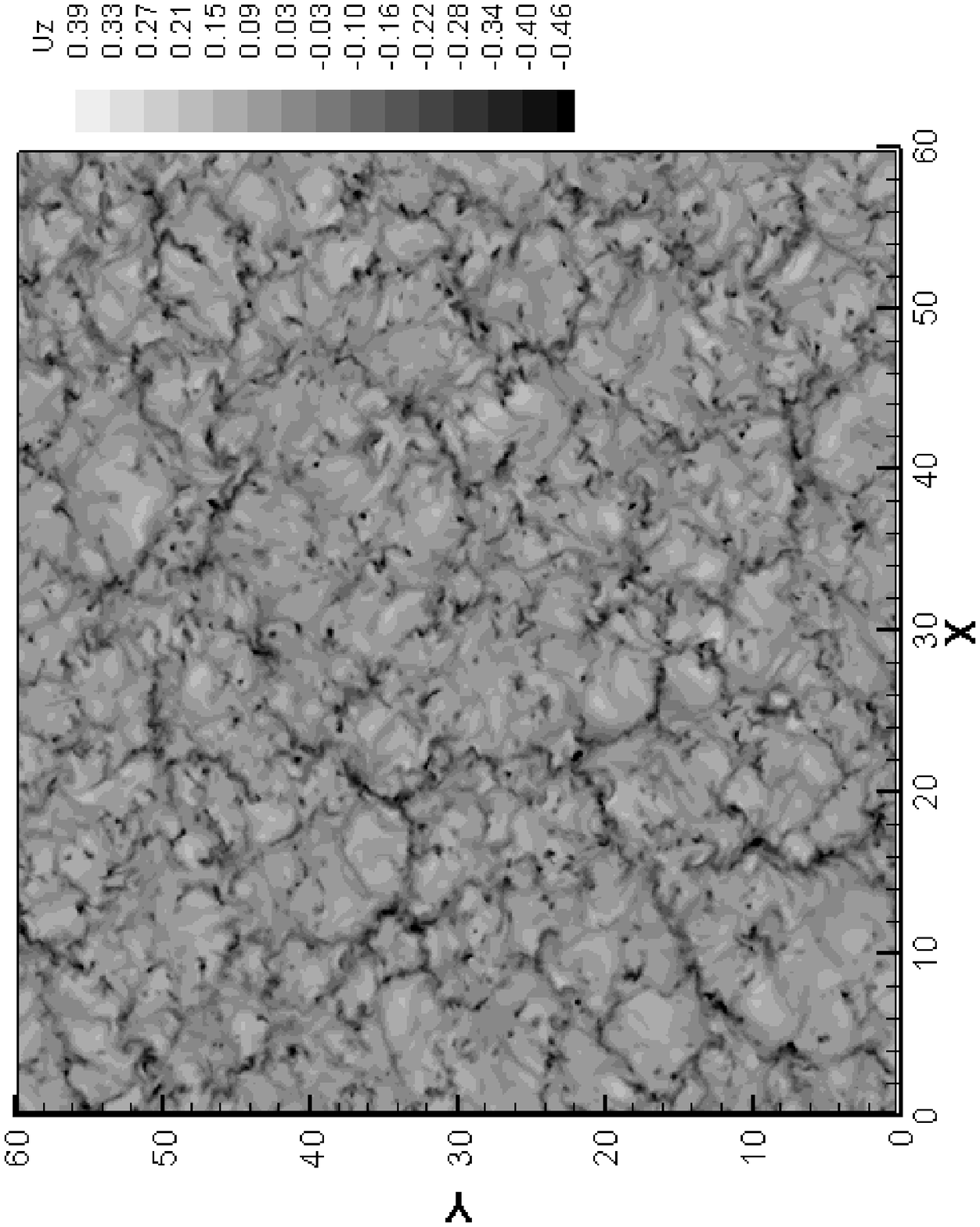}{50mm}{-90}{45}{45}{-200}{180}
  \vspace{2.5cm}
  \caption{\small Contours of the vertical component of velocity at depth 3 Mm. }
\end{figure}

\begin{figure}[!ht]
  \plotfiddle{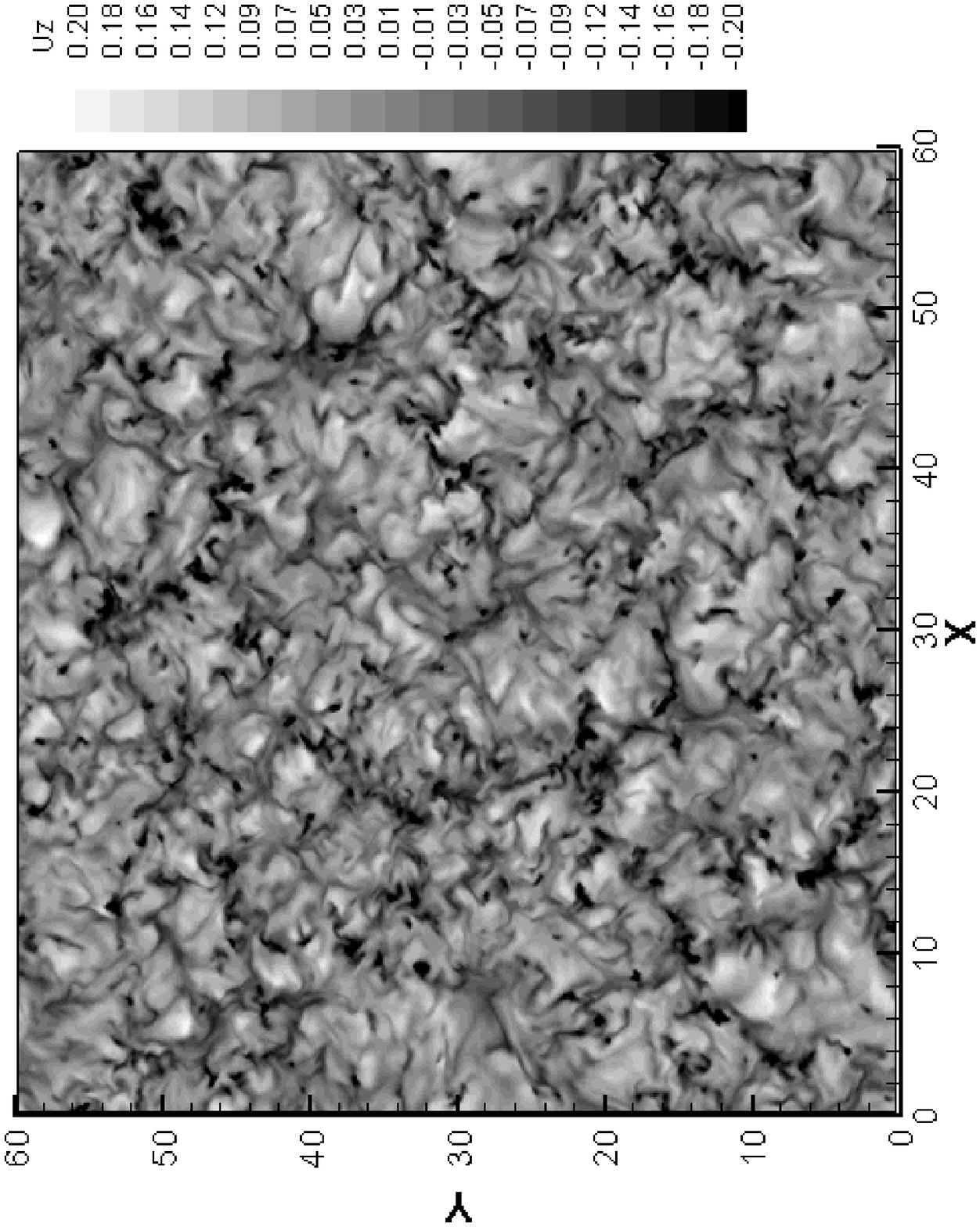}{50mm}{-90}{45}{45}{-200}{180}
  \vspace{2.5cm}
  \caption{\small Contours of the vertical component of velocity at depth 5 Mm. }
\end{figure}

\begin{figure}[!ht]
  \plotfiddle{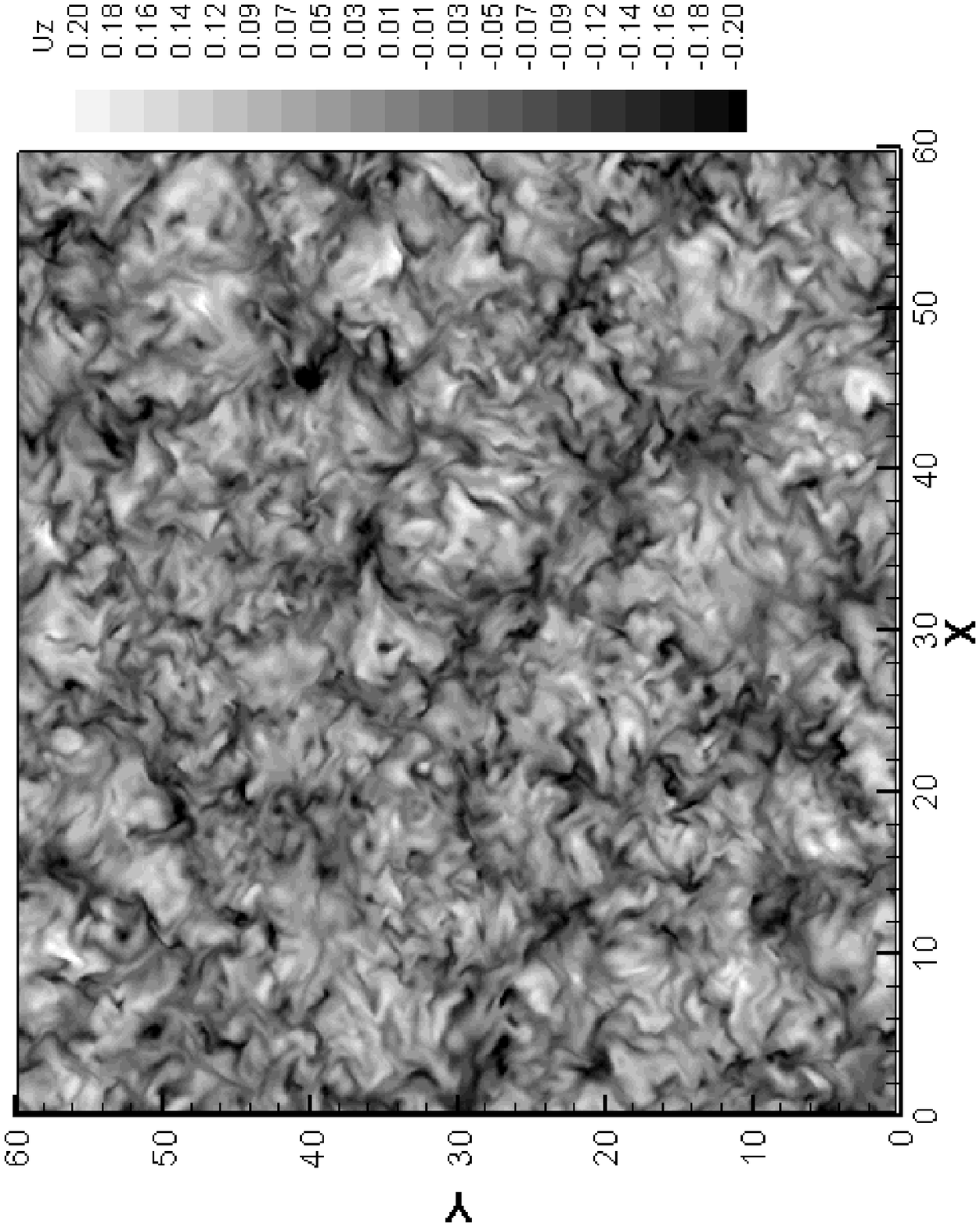}{50mm}{-90}{45}{45}{-200}{180}
  \vspace{2.5cm}
  \caption{\small Contours of the vertical component of velocity at depth 10 Mm. }
\end{figure}

\begin{figure}[!ht]
  \plotfiddle{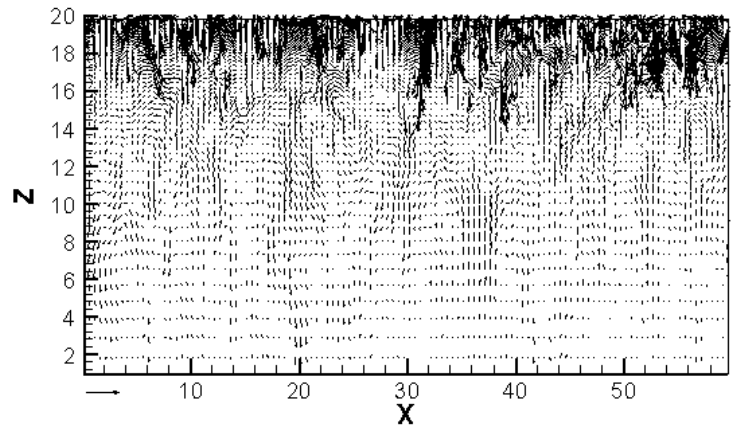}{50mm}{-90}{50}{50}{-200}{180}
  \vspace{3.5cm}
  \caption{\small The velocity field in a vertical plane. }
\end{figure}

\begin{figure}[!ht]
  \plotfiddle{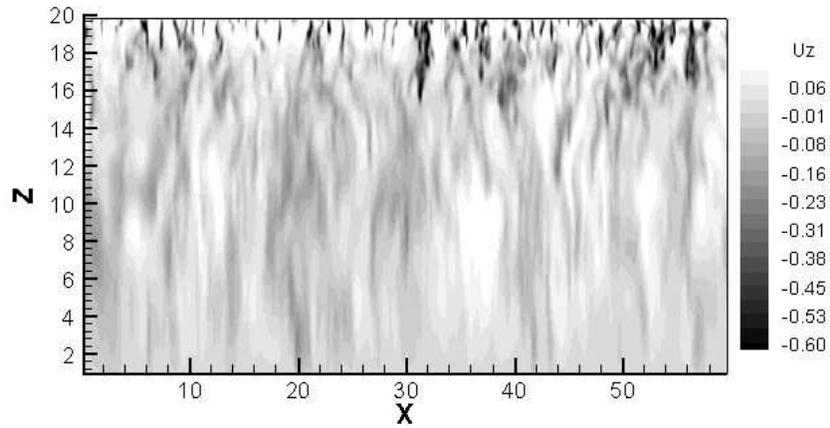}{50mm}{-90}{50}{50}{-200}{190}
  \vspace{3.5cm}
  \caption{\small Contours of vertical component of velocity field. }
\end{figure}

\begin{figure}[!ht]
  \plotfiddle{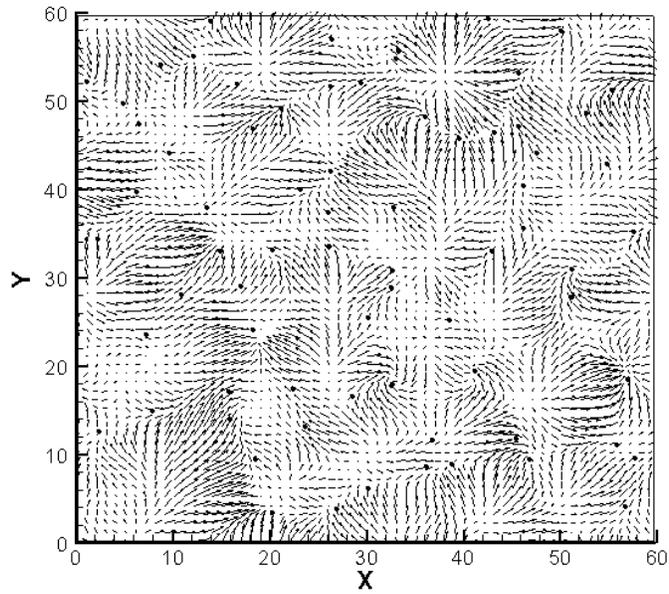}{50mm}{-90}{45}{45}{-200}{180}
  \vspace{3.5cm}
  \caption{\small The velocity field and location of corks. }
\end{figure}

\begin{figure}[!ht]
  \plotfiddle{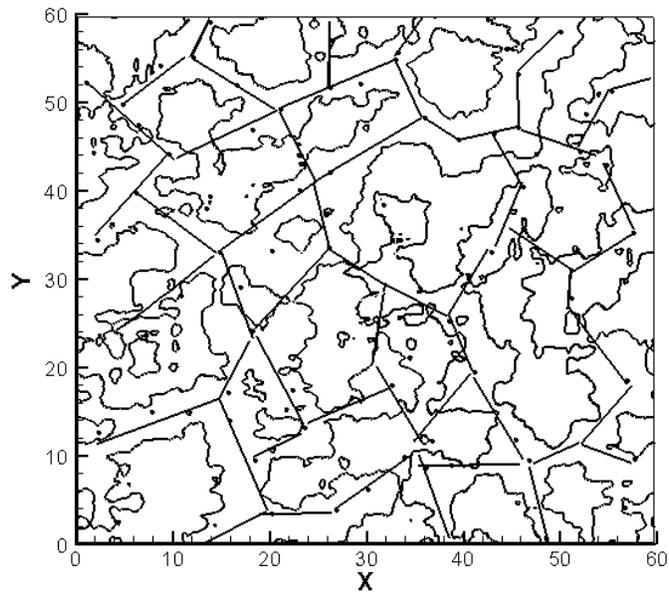}{50mm}{-90}{45}{45}{-200}{180}
  \vspace{2.5cm}
  \caption{\small Null contour of two-dimesional divergence of velocity and boundaries
of supergranulation cells.}
\end{figure}

\end{document}